\documentclass[12pt,preprint]{aastex}

\def\HII{{\ion{H}{2}}}

\def\4959_5007{[\ion{O}{3}]~$\lambda \lambda$4959,5007}
\def\OIII49595007{[\ion{O}{3}]~$\lambda \lambda 4959,5007$}
\def\ratioR23{([\ion{O}{2}]~$\lambda$3727 +
[\ion{O}{3}]~$\lambda\lambda$4959,5007)/H$\beta$}
\def\R23{${\rm R}_{23}$}
\def\dS23{${\rm S}_{23}$}

\def\ratioS23{([\ion{S}{2}]~$\lambda \lambda$6717,31 +
[\ion{S}{3}]~$\lambda\lambda$9069,9532)/H$\beta$}

\def\O4363{[{\ion{O}{3}}]~$\lambda$4363}
\def\OIII{[{\ion{O}{3}}]}

\begin{document}

\title{Internal Dust Correction Factors for Star Formation Rates Derived 
for Dusty \HII\ Regions and Starburst Galaxies}
\author{M. A. Dopita, B. A. Groves \& R. S. Sutherland}
\affil{Research School of Astronomy \& Astrophysics, Australian National
University} 
\authoraddr{Private Bag, Weston Creek PO, ACT 2611, Australia}
\author{L. J. Kewley}
\affil{Harvard-Smithsonian center for Astrophysics, Cambridge, MA}

\begin{abstract}

Star formation rates in galaxies are frequently estimated using the Balmer
line fluxes of their \HII\ regions. However, these can be systematically
underestimated because dust competes for the absorption of Lyman continuum 
photons in the ionized gas. This factor cannot be neglected, and in this paper
we present theoretical correction factors in a simple analytic form in order to
allow observers to take this effect into account when estimating star formation
rates from Balmer lines. These factors scale as the product of the ionization
parameter, ${\cal U}$, and the nebular O/H abundance ratio, both of which can
now be derived from the observation of bright nebular line ratios . 
The correction factors are only somewhat
dependent upon the photoelectron production by grains, but are very sensitive to
the presence of complex PAH-like carbonaceous molecules in the ionized gas,
providing that these can survive in such an environment.

\end{abstract}


\maketitle

\section{Introduction}

\label{Introduction}

A knowledge of the star formation rate (SFR) is fundamental to our
understanding of the formation and evolution of galaxies. In a seminal paper 
\citet{Madau96} connected the star formation in the distant universe with
that estimated from low-redshift surveys, by plotting the estimated star
formation rate per unit co-moving volume against redshift. A wide variety of
techniques have contributed towards populating this diagram with
observational points. Nonetheless, an unacceptable degree of uncertainty
still attaches to our overall understanding of the evolution of star
formation in the universe, and much of this uncertainty results from the
effects of dust obscuration in and around the star-forming regions,
particularly for those techniques which depend upon optical or UV data
\citep[see the review by][]{Calzetti01}.

A fairly direct technique, which has enjoyed extensive use in the 
determination of star formation rates, is the measurement of hydrogen
recombination line fluxes \citep[][ to name but
a few]{Bell01,Gallego95,Jones01, Moorwood00,Tresse98,Pascual01,Yan99,Dopita02b}.
Provided that the \HII\ region can absorb all the EUV photons produced by the
central star, this should be a reliable technique, since the flux in any
hydrogen line is simply proportional to the number of photons produced by
the hot stars, which is proportional to the birthrate of massive stars. This
relationship has been well calibrated at solar metallicity for the H$\alpha $
line \citep{Dopita94,Kennicutt98}: 

\begin{equation}
SFR_{{\rm H\alpha }}=7.9\times 10^{-42}\left[ L_{{\rm H\alpha }}/{\rm erg.s}
^{-1}\right] .  \label{1}
\end{equation}

There exist a number of ways in which this relationship might break down.
First, the \HII\ region might be optically-thin to the EUV
photons in some directions. In this case the recombination flux will provide
an underestimate of the star formation rate, and the appropriate correction
factor is difficult to estimate. Second, there may be dust in front of
the \HII\ region which both reddens and absorbs the recombination lines. In
this case, the corrections for dust absorption can be made using a
foreground screen approximation. This appears to work extraordinarily well
in many galaxies \citep{Bell01, Dopita02b,Kewley02}. Third, dust may be contained
in dense, optically-thick, clouds within or surrounding the \HII\ region. In this
case the optical recombination lines are entirely absorbed in lines of sight
passing through foreground clouds. However, the radio thermal continuum is
transmitted by such clouds, so this factor can be estimated as a ``grey'' screen
by comparison of the optical and the radio data. Such clouds will re-emit the
incident radiation in the far-IR and this emission may also be used to estimate
the stellar EUV flux. 

A fourth possibility is that dust within the ionized gas itself competes
with the gas to absorb the EUV photons from the central star(s). This
possibility, first seriously quantified by \citet{Petrosian72}, has been
discussed by a  number of authors since \citep{Panagia74, Mezger74,
Natta76, Sarazin77, Smith78, Shields95, Bottorff98}.
More recently its effect has investigated and quantified (as far as possible
by direct observation) in a series of recent papers by Inoue and his
collaborators \citep{Inoue00, Inoue01a, Inoue01b}. Here we attempt to calibrate
these correction factors using theoretical models of dusty \HII\ regions.

\section{The Models}
\label{models}

\subsection{The Dust Model}

The MacOS version of the MAPPINGS IIId code was used to generate
spherical-shell dusty photoionized models appropriate to \HII\ regions excited by
clusters of OB stars. MAPPINGS IIId includes a number of advances in
the treatment of dust physics and absorption. The dust model includes three
types of grains, ``astronomical'' silicates and graphite or amorphous carbon
grains having a \citet{MRN77} (MRN) distribution for grain
sizes between 50\AA\ and 2500\AA . For these grains the extinction curve is
based on the \citet{Laor93} data for silica and graphite. In
addition, we have the option of including complex PAH-like organic
molecules. For these, the photoionization cross-section per carbon atom
follow the yield factors given by \citet{Vestraete90} which are based
on pyrene and coronene. These have been extrapolated (using the same slope as
the Vestraete curve) to $N_{{\rm c}}=80$, which seems to be appropriate for
interstellar PAHs \citep{Allain96a, Allain96b, Hollenbach99, Peeters02}. For PAH
molecules of these dimensions, the ionization potential is about 6~eV. The PAH
opacities are taken from the recent publication by \cite{Li02}.

With our present state of knowledge on the physics of PAH molecules, it is
not at all clear whether such molecules can survive for significant lengths
of time in the hostile environment offered by the ionized zones of an \HII\
region. Provided that the heating rate by the absorption of photons in the
ISM can be at least matched by the infrared radiative rate, then the survival of
polycyclic aromatic hydrocarbons, PAHs, is set by the competition between
photodissociation (by the ejection of an acetyleneic group) and its repair
through accretion of carbon atoms \citep{Allain96a,Allain96b}. If $\tau _{{\rm
diss}}$ is the radiative dissociation timescale, and $\tau _{{\rm acc}}$ is
the C atom accretion timescale, then these are given by 

\begin{equation}
\tau _{{\rm diss}}=\left( F_{{\rm FUV}}\sigma _{{\rm diss}}\right) ^{-1},
\label{2}
\end{equation}
and 
\begin{equation}
\tau _{{\rm acc}}=\left( n_{{\rm H}}X_{{\rm C}}k_{{\rm acc}}\right) ^{-1},
\label{3}
\end{equation}
where $F_{{\rm FUV}}$ is the far-UV radiation field, $\sigma _{{\rm diss}}$
is the photodissociation cross section per PAH molecule, $n_{{\rm H}}$ is
the number density of hydrogen atoms, $X_{{\rm C}}$ is the abundance of C in
the ISM, and $k_{{\rm acc}}$ is the reaction rate for sticking of a carbon
atom onto such a molecule. For the far-UV radiation field, we take the total
radiation field above the ionization potential adopted for these molecules ( 
$\ge 6$~eV). Putting together equations \ref{1} and \ref{2}, we see that the
PAHs will be destroyed when 
\[
F_{{\rm FUV}}\sigma _{{\rm diss}}>n_{{\rm H}}X_{{\rm C}}k_{{\rm acc}},
\]
or 
\begin{equation}
{\cal H}=\frac{F_{{\rm FUV}}}{cn_{{\rm H}}}>\frac{X_{{\rm C}}k_{{\rm
acc}}}{\sigma _{{\rm diss}}}.  \label{4}
\end{equation}
We shall define ${\cal H}$ as the {\em Habing Photodissociation
Parameter}, by analogy with the dimensionless Ionization Parameter ${\cal U}$
used in \HII\ region theory. The advantage of the use of this parameter is
that all photodissociation rates will scale in this way, and therefore 
the local value of this dimensionless parameter will also determine the local
chemistry of photodissociation regions to first order. The actual shape of
the photodissociating spectrum will determine the chemistry to second order.

Since we do not know the absolute value of ${\cal H}$ above which PAHs are
destroyed, we have to ask, what are the most extreme values of ${\cal H}$
observed in regions which still contain PAHs. From Allain et al. (1996a), such
extreme  regions can be identified as in the diffuse ISM, high above
the galactic plane where $n_{{\rm H}}\sim 0.1$~cm$^{-3}$, $F_{{\rm FUV}}\sim
1.5\times 10^{8}$ photons cm$^{-2}$ s$^{-1}$ or in the planetary nebula NGC7027
where $n_{{\rm H}}\sim 7\times 10^{4}$cm$^{-3}$ and $F_{{\rm FUV}}\sim 7.6\times
10^{13}$ photons cm$^{-2}$ s$^{-1}$, corresponding to ${\cal H}\sim 0.05$ and
${\cal H}\sim 0.04$ respectively. We therefore adopt a threshold of ${\cal
H}\sim 0.05$ for the destruction of PAHs. This threshold is not reached for any
of the photoionization models presented here, so it is at least conceivable that
charged PAH-like molecules may survive in the environment of an \HII\ region.

\subsection{The Photoionization Models}

The photoionization modelling procedure adopted here closely follows that
described in \citet{Dopita00} and \citet{Kewley01}. For the central star cluster,
we adopt a STARBURST99 \citep{Leitherer99} instantaneous-burst with a total
luminosity of $10^{40}$~erg~s$^{-1}$ and having an Initial Mass Function (IMF)
with a power-law slope of the Salpeter form ($\alpha =2.35$). The lower mass
cutoff was set at 0.1$M_{\odot }$ and the upper mass cutoff at 120 $M_{\odot }$.
For the purpose of the models presented  here, the actual form of the
assumed IMF is relatively unimportant in determining the final results,
since the strength of the EUV field and the metallicity of the gas prove to
be more important than the shape of the ionizing spectrum in determining the
EUV absorption by dust. Table 1 gives the abundance set used in the models,
and our assumed gas-phase depletion factors. These differ slightly from the
Dopita \& Kewley (2000) models because we have adopted the revised solar
abundances for O, C, Si and Fe \citep{Asplund00a, Asplund00b, Allende01,
Allende02}. 

In order to investigate the effects of ionization parameter, ${\cal U}$, and
the chemical abundances (which effectively determines the gas-to-dust
ratio), we ran three sets of models for the abundances $Z=0.4, 1.0$~and
$2.0Z_{\odot}$, and covering a range of initial dimensionless ionization
parameter from ${\cal U} = 0.000$ up to ${\cal U} = 0.0066$. This encompasses 
the full range that is normally encountered in
range of ionization parameters encountered in bright \HII\ regions (see
Dopita \& Kewley, 2000). Each model had the chemical abundances of the central
cluster set the same as for the gas in the \HII\ region.  In spherical \HII\
region models, the spherical divergence of the radiation field ensures that the
mean ionization parameter is not well defined when the \HII\ region becomes
thick in comparison to its radius. We have therefore ensured that the models
remain geometrically thin with a well-defined ${\cal U}$ by raising the assumed
hydrogen density from $n = 10$~cm$^{-3}$ up to $n = 100$~cm$^{-3}$ for the
models with the highest ionization parameter.

The absorption of the PAH molecules is very important in determining the EUV
extinction, since they have a very large absorption cross-section per
carbon atom above 13.6~eV. Since we have no way of directly determining
whether such molecules survive in the \HII\ region itself, we have run
two sets of models, one which did not include PAH-like molecules, and
another in which we have set the abundance of PAH molecules equivalent to 20\%
of the total carbon atoms. This is probably an overestimate of their true
abundance, since only a maximum of 40\% of interstellar carbon is locked up in
carbonaceous grains \citep{Duley99}, and studies of PAH emission suggests that
perhaps only about 10\% of the interstellar carbon is actually locked up in
PAH-like molecules \citep{ Li02}.

For most models, we have also run a comparison dust-free \HII\ region model with
the same gas-phase abundances as in our dusty models in order both to check the
absolute value of the dust-free H$\beta$ recombination line flux, and to ensure
that this is independent of the ionization parameter for a given input
spectrum, as required by theory. We find that, over the metallicity range
covered by these models, the EUV blanketing of the central star cluster has very
little effect on the number of Lyman continuum photons produced per unit of
stellar luminosity (less than 0.02~dex).

For reasons fully explained in the next section, we expect that the fraction of
ionizing photons absorbed by the dust should scale as the product ${\cal
U}({\rm O}/{\rm H})$, for a given grain model. Therefore in Figure (1) we have
plotted this quantity against the computed ratio of Balmer H${\beta}$ fluxes,
with and without the inclusion of dust. The two families of curves are shown,
one with 10\% of the C locked in PAH-like molecules, and the other with 20\% of
the C locked in such molecules. The families of models with at 0.4 times solar
abundance (open circles), 1.0 times solar abundance (crossed circles) and 2
times solar abundance (filled circles) lie close to each other for each grain
composition. This gives us confidence that the ${\cal U}({\rm O}/{\rm H})$
scaling factor is correct, to first order.

The largest dust absorption is found for the models with extreme
metallicity and ionisation parameter (${\cal U}=0.0133$ and $Z=2Z_{\odot}$). In
practice \HII\ regions are rarely encountered with such a high ionization
parameter \citep{Dopita00, Kewley02}. Therefore, we conclude on the basis of our
models that accounting for the effect of dust absorption in determining star
formation rates is both a significant and important correction, but that dust
rarely dominates in the absorption of the EUV photons in normal \HII\ regions.

\subsection{An Simple Analytic Fit}

It is self-evident that greater dust content will enable the dust to compete
more efficiently with the gas for the EUV Lyman continuum photons. The dust 
content is determined by the balance of grain formation and grain destruction
processes, but it should, to first order, scale as the metallicity, $Z$. We
have taken this as an assumption of our models by using the same set of
depletion factors in all models. However, it is not so immediately apparent
that for high ionization parameter, the dust becomes again relatively more
important in the competition for absorption of the ionizing  photons. This result
is readily established. After Dopita et al. (2002a), the local absorption of
ionizing photons by the ionized plasma is simply equal to the local
recombination rate:

\begin{equation}
\frac{dS_{*}}{dx}=-\alpha \left( T_{e}\right) n^{2},  \label{5}
\end{equation}
where $n$ is the density in the ionized plasma, $S_{*}$ $=\left( hc\right)
^{-1}\int \lambda I\left( \lambda \right) d\lambda $ is the local photon density
from the ionizing source (cm$^{-2}$s$^{-1}$), and $\alpha \left(
T_{e}\right) $ is the recombination coefficient. Here, we have implicitly
assumed that the nebula is fully ionized, and that $n=n_e=n_{\rm H}$. The
absorption of photons by dust is given by 

\begin{equation}
\frac{dS_{*}}{dx}=-\kappa nS_{*},  \label{6}
\end{equation}
where $\kappa $ is the effective dust opacity (per atom). 
It follows that dust absorption becomes relatively more important as the strength
of the ionizing field increases and dominates the absorption of photons in the
photoionized plasma when

\begin{equation}
{\cal U}>\frac{\alpha \left( T_{e}\right) }{c\kappa }.  \label{7}
\end{equation}
Here ${\cal U} = S_{*}/cn$ is the dimensionless ionization parameter and $c$
is the speed of light. Substituting numerical values appropriate for solar
abundance, we find that the critical ionization parameter above which dust
dominates the absorption is $\sim $0.01. This is in good agreement with what is
indicated by detailed modelling. 

In fact, we can quite readily compute the fraction of ionizing photons absorbed
by a dusty \HII\ region model relative to a dust-free model. Dropping the
explicit dependence of $\alpha$ on $T_e$, the radiative transfer equation for a
plane-parallel (geometrically thin) nebula including both the gas (equation
\ref{5}) and the dust (equation
\ref{6}) terms is

\begin{equation}
\frac{dS_{*}}{dx}=-\alpha \left( T_{e}\right) n^{2} - \kappa nS_{*}. \label{8}
\end{equation}

Solving this in terms of the ionization parameter at the inner edge of the
nebula  ${\cal U}_{0}$;

\begin{equation}
\ln{\left[{\frac{\alpha}{c\kappa }}\right]} -\ln{\left[{\frac{\alpha}{c\kappa }}
+  {\cal U}_{0}\right]}= {x_{d}\kappa n} = {{\tau}_{d}},  \label{9}
\end{equation}
where $x_{d}$ is the thickness of the ionized layer (with dust), and
${\tau}_{d}$ is the optical depth in dust through the ionized layer. In the
absence of dust, we can integrate equation \ref{5} to solve for the thickness
of the ionised layer; $ x_{o}= c{\cal U}_{0}/{\alpha}n$. Therefore, in a nebula
of uniform density, the ratio of the recombination line flux with and without
dust is simply $f=F_{{\rm H}\beta}/F_{{\rm H}\beta}(0)=x_d/x_o$. Thus it follows
from equation \ref{9} that

\begin{equation}
{f={\frac{F_{{\rm H}\beta}}{F_{{\rm
H}\beta}(0)}=y^{-1}\ln \left[\frac{1}{1+y}\right]}},  \label{10}
\end{equation}
where $y=(c/{\alpha}){\cal U}_{0}\kappa$. This fraction has the same meaning as
the fraction $f$ defined by \citet{Inoue01a} or \citet{Petrosian72} for
spherical filled \HII\ regions.

Since the dust opacity, $\kappa$, scales as the metallicity ({\emph i.e.} as
the O/H abundance by number of atoms) it follows that that the fraction of EUV
photons absorbed by the dust is a function only of the product of the
metallicity, $Z$, or (equivalently) (O/H), and the initial ionization parameter,
${\cal U}_0$. This scaling is the physical reason why \citet{Inoue01a} found an
inverse correlation between not only $f$ and log(O/H), but also an inverse
correlation between $f$ and the number of Lyman photons estimated from the
central star(s). It is also significant that the \HII\ regions having the most
luminous clusters also tend to have low densities, consistent with them having
high values of ${\cal U}_{0}$. These \HII\ regions also tend to have the lowest
inferred values of $f$.

Equation \ref{10} is strictly valid if the effect of absorption of EUV
photons on dust is only to remove photons from the EUV field. However, grains
can also produce photoelectrons, and some of these may recombine with a proton
to produce a additional contribution to the Balmer emission. Provided that only a
small percentage of dust absorptions produce a photoelectron, then to first
order the corrected absorption fraction$f^{'}$, is 

\begin{equation}
f^{'}=(1-f)\rm Y , \label{11}
\end{equation}
where $\rm Y$ is the photoelectric yield, of order 0.1. The importance of
photoelectric emission is suggested by the fact that, on figure 1, the models
having twice solar abundance lie systematically higher than the models with
lower abundance. This is the result of two effects. First, the higher dust to
gas ratio ensures that there are more dust-produced photoelectrons per unit mass
of gas. Second, at high heavy element abundance, the electron temperature of the
\HII\ region is much lower, and the recombination coefficient of hydrogen is
therefore higher, ensuring a greater production of Balmer photons by
recombining dust-produced photoelectrons.

The dust absorption curve without photoelectric emission (equation \ref{10})
and with photoelectric emission ($Y=0.1$) (equation \ref{11}) were scaled to fit
the two families of models (with 0\% and 20\% of C in PAHs, repectively).
These two families of curves probably represent the extremes allowed by the
theoretical models. For the 0\% PAH models, $y=1$ corresponds to
$\log[{\cal U}_{0}\times (O/H)]=-5.377$, while for the 20\% of C in PAH models,
$y=1$ is reached by $\log[{\cal U}_{0}\times {\rm O/H}]=-5.854$. This emphasizes
how efficient PAHs (if present) are in increasing the opacity of the \HII\
region to dust. This is because PAHs offer an essentially a molecular opacity,
with the carbon atoms arranged in flat sheets so that relatively few atoms can
offer a large cross-section to the EUV radiation field.

The dust absorption in spherical models is more severe than for the thin-shell
models presented here. \citet{Inoue01a} and \citet{Petrosian72} showed that, for
such models,  $f=F_{{\rm H}\beta}/F_{{\rm
H}\beta}(0)={{\tau}_d^3/3[({\tau}_d^2-2{\tau}_d}+2)\exp({\tau}_d)-2]$. If the
initial slope of this function is fitted to our thin shell \HII\ region models,
it rapidly becomes too steep at larger $\log[{\cal U}_{0}\times ({\rm O/H})$, and
$f$ approaches zero. The reason for this is that the inner regions of filled
spherical \HII\ regions are characterized by a very high local radiation field,
so that dust may compete much more efficiently for EUV photons in this region.
Thus, overall, the filled \HII\ regions will provide an upper limit to the dust
absorption fraction. However, since the geometry of most extragalactic \HII\
regions is better fit with a central empty zone, such an empty zone is naturally
produced by the strong stellar winds of the OB stars in \HII\ regions, and since
the strong-line diagnostics of extragalactic \HII\ regions are also well fit by
shell models \citep{Dopita00, Kewley02}, we are of the opinion that spherical
shell models should be more physically realistic than filled Str\"omgren spheres
in most cases.

\begin{figure}
\plotone{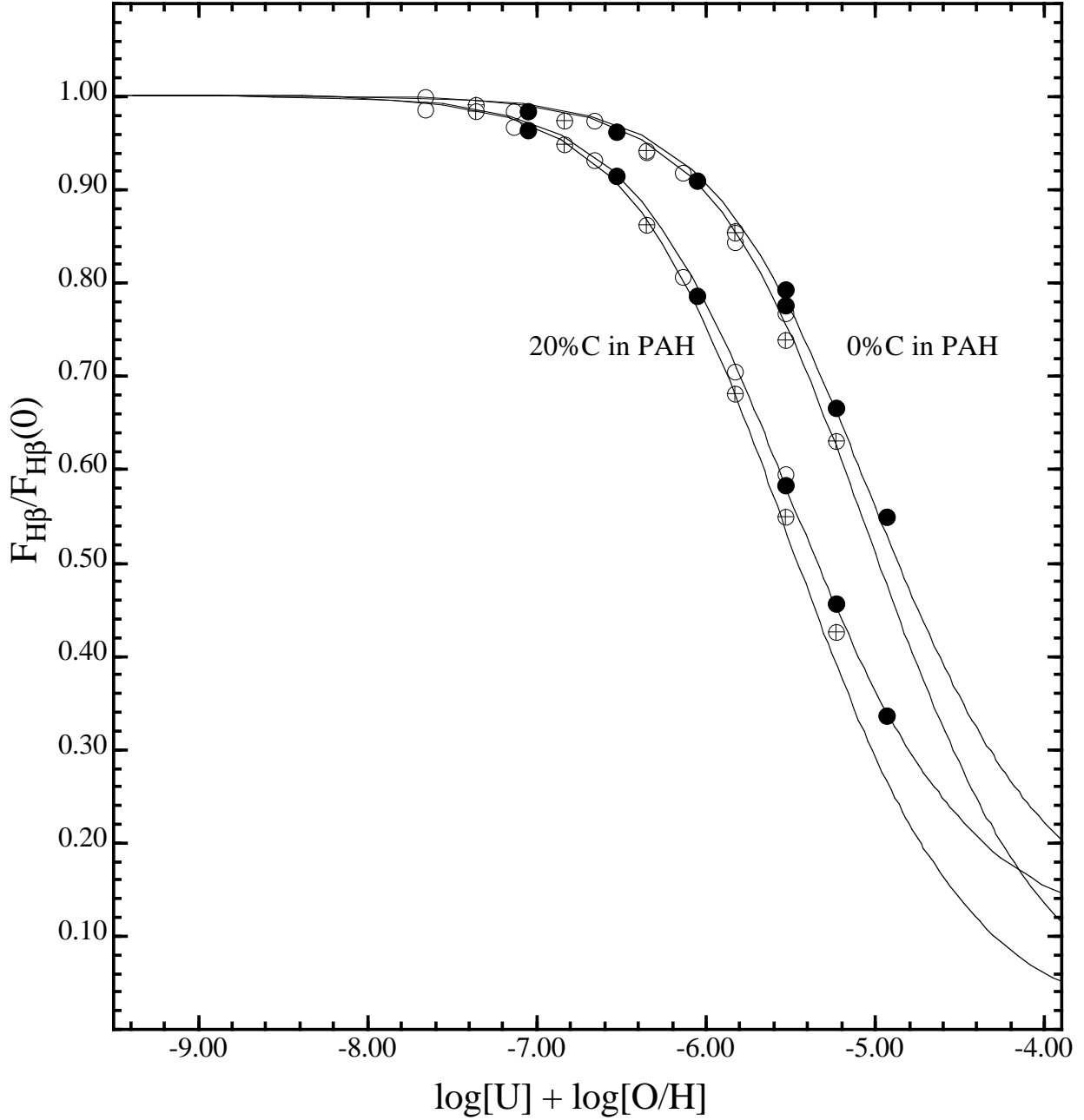}
\caption[fig_1.eps]{The computed fraction of the potential H$\beta$ flux 
emerging from dusty model \HII\ regions as a function of the product of initial
ionization parameter ${\cal U}$ and O/H abundance by number. The circles are the
models; filled circles representing twice solar abundances, the crossed circles
solar abundance, and the open circles, 0.4 times solar abundance. The upper set
of models is for a amorphous carbon/ graphite/ astronomical silicate grain mix.
In the lower set, the PAH-like organic molecules account for 20\% of
the total carbon abundance (an upper limit to the likely range of values). For
each set, the analytic solution described in the text has been fit to the data.
In each pair the lower curve is without a photoelectric contribution, and the
upper curve is for a 10\% photoelectric yield. 
\label{fig_1}}
\end{figure}

\placefigure{fig_1}

\section{Conclusions}

We have studied the theoretical effect of internal dust in absorbing the Lyman
continuum photons, and so in reducing the Balmer line fluxes produced by \HII\
regions. We agree with the conclusion of \citet{Inoue01a} that ``the effect of
Lyman continuum extinction is not negligible relative to other uncertainties of
estimating the star formation rates of galaxies''. However, the largest
uncertaintly in correcting the estimated star formation rates for internal dust
absorption is the presence or absence of PAH-like organic molecules in the
ionized gas. Since this is so theoretically uncertain, we have run models either
without these molecules, or else including these at the level of 20\% of total
carbon, which is probably a reasonable upper limit for their abundance.

We have also provided simple theoretical fits to the fraction of the EUV photons
which are used by the ionized gas, $f$. On the assumption that the dust to gas
ratio scales as the metallicity of the gas, this fraction $f$ is shown to be
only a function of the product of the initial ionization parameter, ${\cal U}$
and the oxygen abundance (O/H). This is fortunate, because each of these can
be estimated independently for a given \HII\ region or starburst galaxy by
measuring only the strong emission lines at optical wavelengths (ideally the
[{\ion{O}{2}}], H$\beta$, [{\ion{O}{3}}], H$\alpha$, [{\ion{N}{2}}] and
[{\ion{S}{2}}] lines, although subsets of these may be used in
certain circumstances \citep{Kewley02}). 

Once both ${\cal U}$ and (O/H) have been determined, the $f-$factor can then be
obtained by either reading directly from figure 1, or else by solving for $f$
using the analytic fits given by equations \ref{10} and \ref{11}, with $y=1$
corresponding to $-5.854 < \log[{\cal U}_{0}\times {\rm O/H}] < -5.377$. Within
these theoretical uncertainties, the star formation can then be derived using the
corrected \citet{Kennicutt98} expression; ${\rm SFR}_{{\rm H\alpha }}=7.9\times
10^{-42}\left[ L_{{\rm H\alpha }}/{\rm erg.s} ^{-1}\right]/f$. We could also use 
an analagous expression involving any other member of the Balmer, Paschen or 
Brackett Series of hydrogen.

\section*{Acknowledgements}

M. Dopita acknowledges the support of the Australian National University and 
the Australian Research Council through his ARC Australian Federation Fellowship, 
and under the ARC Discovery project DP0208445. M. Dopita would also like to thank 
United Airlines for the provision of airport lounges which gave him the 
opportunity to run most of the models described here on his laptop computer. 
L. Kewley is supported by a  Harvard-Smithsonian CfA Fellowship.

\clearpage

\begin{deluxetable}{lcc} 
\tabletypesize{\small}
\tablewidth{3in}
\tablecaption{Solar metallicity ($Z_{\odot}$) and gas-phase depletion
factors (D) adopted for each element.\label{Z_table}}
\tablehead{
\colhead{Element} 
& \colhead{$\log({\rm Z_{\odot}})$}
& \colhead{$\log({\rm D})$}\\
}
\startdata
H & 0 & 0.00 \\
He & -1.01 & 0.00 \\ 
C & -3.44 & -0.30 \\
N & -3.95 & -0.22 \\
O & -3.35 & -0.07 \\
Ne & -3.91 & 0.00 \\
Cl & -5.75 & 0.00 \\
Mg & -4.42 & -0.70 \\
Si & -4.45 & -1.00 \\
S & -4.79 & 0.00 \\
Ar & -5.44 & 0.00 \\
Ca & -5.64 & -2.52 \\
Fe & -4.36 & -2.00 \\
Ni & -5.68 & -2.00 \\
\enddata
\end{deluxetable}
\normalsize

\newpage

\end{document}